\documentclass[%
aip,jap,letterpaper,floatfix,
 preprintnumbers,
 reprint,
 amsmath,amssymb,
 showpacs,showkeys,
 superscriptaddress,
]{revtex4-1}
\usepackage{lineno,xcolor}
\usepackage{graphicx}
\usepackage{bm}
\usepackage[breaklinks=true]{hyperref}
\hypersetup{bookmarksnumbered, pdfpagemode=UseOutlines, 
pdfauthor={A.\ Aqeel}, 
pdftitle={Surface sensitivity of spin Seebeck effect}, pdfdisplaydoctitle, 
colorlinks=true, citecolor=blue, filecolor=blue, linkcolor=blue, urlcolor=blue}
\usepackage{enumitem}
\usepackage{epstopdf}
\usepackage{pdfpages}

\date{\today}
\begin{document}


\title{Surface sensitivity of the spin Seebeck effect}
\author{A.\ \surname{Aqeel}}
\author{I.\ J.\ \surname{Vera-Marun}}
\author{B.\ J.\ \surname{van Wees}}
\author{T.\ T.\ M.\ \surname{Palstra}}
	\email[e-mail: ]{t.t.m.palstra@rug.nl}
\affiliation{Zernike Institute for Advanced Materials, University of Groningen, Nijenborgh 4, 9747 AG Groningen, The Netherlands}

\begin{abstract}
We have investigated the influence of the interface quality on the spin Seebeck effect (SSE) of the bilayer system yttrium iron garnet (YIG) $-$ platinum (Pt). The magnitude and shape of the SSE is strongly influenced by mechanical treatment of the YIG single crystal surface. We observe that the saturation magnetic field ($H_{\text{sat}}^{\text{SSE}}$) for the SSE signal  increases from 55.3 mT to 72.8 mT with mechanical treatment.  The change in the magnitude of $H_{\text{sat}}^{\text{SSE}}$ can be attributed to the presence of a perpendicular magnetic anisotropy due to the treatment induced surface strain or shape anisotropy in the Pt/YIG system. Our results show that the SSE is a powerful tool to investigate magnetic anisotropy at the interface.

\end{abstract}

\keywords{}

\maketitle
\section{Introduction}
The discovery of the spin Seebeck effect (SSE) \cite{Uchida_nmat_2010} in insulators triggered the modern era of the field of spin caloritronics \cite{bauer_spincaloric_2012}. In insulators, instead of moving charges, only spin excitations (magnons) drive the non-equilibrium spin currents. In the spin Seebeck effect, spin currents are thermally excited in a ferromagnet FM and detected in a normal metal NM deposited on the FM. The bilayer NM/FM system in the SSE provides the opportunity to separately tune the properties of both layers to optimize the magnitude and magnetic field dependence of the SSE effect. The platinum (Pt) and yttrium iron garnet (YIG) bilayer system has attracted considerable attention for studying the spin Seebeck effect \cite{Uchida_nmat_2010,uchida2010,kehlberger2013,kirihara2012} and for other spin dependent transport experiments \cite{Kajiwara2010,Burrowes2012,Uchida_nmat_2010,Sandweg2011,Heinrich2011,Rezende2013,Qui2012,Vlietstra_spinmix2013,flipse_observation_2013}. Platinum (Pt) has a large inverse spin Hall response  \cite{ishidaTM2013} whereas YIG is an ideal ferromagnetic insulator due to low magnetic damping \cite{bauer_spincaloric_2012,Kajiwara2010,Chikazumi_book} and a large band gap \cite{JiaSTT2011} at room temperature.  

The origin of the spin Seebeck effect is commonly explained by the difference in the magnon temperature in the FM and the phonon temperature in the NM, $\Delta T_{mp}$  \cite{Xiao2010,Adachi2013}. When the temperature gradient  $\nabla T$ is applied across the NM/FM system, it creates a $\Delta T_{mp}$ based on the thermal conductivities of the magnon and the phonon subsystems \cite{Xiao2010}. This $\Delta T_{mp}$ induces a spin current density at the interface which is detected in the normal metal Pt by the inverse spin Hall effect (ISHE). The ISHE signal depends on a scaling parameter, the interfacial SSE coefficient $L_s$, related to how efficiently the spin current density can be created across the interface under a certain $\Delta T_{mp}$. The resulting spin Seebeck signal scales linearly with the length of the NM ($l_{Pt}$), therefore for the Pt/YIG system

\begin{equation} \label{eq:V_ISHE}
V_{\text{ISHE}} \propto l_{Pt} \ . \ L_s \ . \ \nabla T
\end{equation}          

The scaling parameter $L_s$ is proportional to the real part of the spin mixing conductance $g_r ^{\uparrow \downarrow}$ at the interface. The spin mixing conductance $g_r ^{\uparrow \downarrow}$ and therefore the SSE are very sensitive to the interface quality \cite{jungfleisch2013}. In recent years substantial effort has been made to improve the spin mixing conductance on thin films of YIG \cite{qiu2013,jungfleisch2013} and bulk crystals  \cite{JiaSTT2011,UchidaFerrites2013}. Unlike thin films, bulk crystals need an extra surface polishing step for the device fabrication, due to the initial surface roughness. The polishing of the crystal surface can influence the spin mixing conductance in several ways. Apart from changing the surface roughness, mechanical polishing can change the magnetic structures at the interface by inducing a small perpendicular anisotropy at the surface layer of the YIG crystal \cite{craik1959,craik1961,coleman1974}.  However, the effect of polishing on the spin Seebeck effect (SSE) has not yet been systematically studied. In this paper, we report the effect of mechanical surface treatment of the YIG single crystals on the SSE effect. This systematic study reveals the surface sensitivity of the SSE and indicates new ways of surface modification for improved thermoelectric efficiency. 
\section{Experimental technique}
\begin{figure}[tbp]
 \centering
\includegraphics[width=0.5\textwidth,natwidth=310,natheight=342]{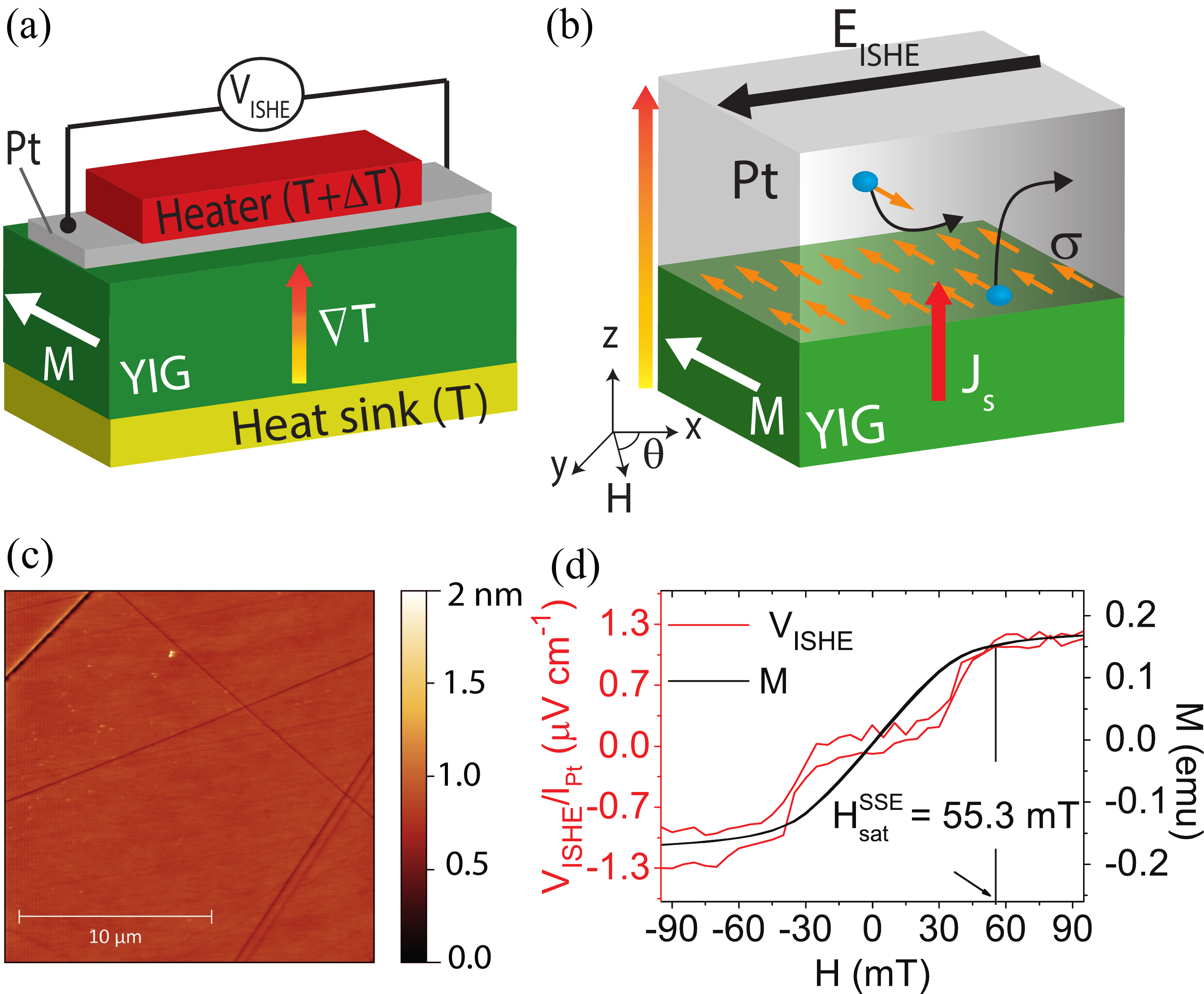}
\caption{\label{fig:5}
(a) Device configuration of the longitudinal SSE where $\nabla T$ represents the temperature gradient across the Pt/YIG system. (b) Detection of spin current by the ISHE. The orange arrows indicate the spin polarization $\sigma$ at the interface of the Pt/YIG system. M, $J_S$ and $E_{\text{ISHE}}$ represent the magnetization of YIG, spatial direction of the thermally generated spin current, and electric field induced by the ISHE, respectively. $\theta$ represents the angle between the external magnetic field H in the x-y plane and the x axis. (c) AFM height image of a single crystal YIG surface (20 x 20 $\mu m^2$) for sample S1. (d) a comparison between the magnetic field dependence of $V_{\text{ISHE}}$  at $\Delta T$ = 3.6 K for sample S1 and the magnetization M of the YIG crystal.}
\end{figure}

In this study, we use the longitudinal configuration \cite{uchida2010} for the spin Seebeck effect where the temperature gradient is applied across a NM/FM interface and parallel to the spin current direction $J_s$. In Fig.~\ref{fig:5}(a), we illustrate schematically the device configuration for measuring the SSE used in this study. The sample consists of a single crystal YIG slab and a Pt film sputtered on a (111) surface of the YIG crystal.  When an out-of-plane (along z-axis) temperature gradient is applied to the Pt/YIG stack, spin waves are thermally excited. The spin waves inject a spin current along the z-axis and polarize the spins in  the Pt film close to the interface parallel to the magnetization of the YIG crystal, as illustrated in Fig.~\ref{fig:5}(b). Due to the strong spin-orbit coupling in the Pt-film, the spin polarization $\sigma$ is converted to an electrical voltage $V_{\text{ISHE}}$.
The single crystals of YIG with the same purity were used in all measurements. The YIG crystals were grown by the floating zone method along the (111) direction and commercially available from Crystal Systems Corporation company, Hokuto, Yamanashi Japan . A diamond saw was used to cut the crystals. The YIG crystals  were cleaned ultrasonically first in acetone and then ethanol baths.

Three different types of surfaces were prepared for samples S1~-~S3 by the following treatments:
\begin{itemize}[noitemsep,leftmargin=*]
\item For S1: the YIG crystals were grinded with abrasive grinding papers (SiC P1200 - SiC P4000) at 150 rpm for 1h. After grinding, diamond particles were used with a sequence of 9 $\mu m$, 3 $\mu m$ and 1 $\mu m$ at 300 rpm for 30 mins, respectively. To remove the surface strain or surface damage due to diamond particles  \cite{craik1959,craik1961,coleman1974}, colloidal silica OPS (oxide polishing suspension) with a particle size of 40 nm was used, which can give mechanical as well as chemical polishing. To remove the residuals of polishing particles, samples were heated at 200 $^\circ$C for 1h at ambient conditions. Then crystals were cleaned by acetone and ethanol in an ultrasonics bath before depositing the Pt layer on top. 
\item For S2: grinding, polishing and cleaning of the samples were done in the same way as described for S1. However, the colloidal silica OPS was not used for sample S2. Thus, the strained or damaged surface layer due to diamond polishing was retained. 
\item For S3: no mechanical polishing was done to obtain flat surfaces as done for samples S1 and S2. After cleaning in the same way as done for samples S1 and S2, Pt was deposited on the unpolished YIG crystal surface.  
\end{itemize}
\begin{table}[ht]
\small
\caption{Surface treatment, surface roughness, and orientation of the YIG crystals for different samples.}
\centering
\begin{tabular}{c c c c c c c}
\hline
\hline
Samples & Polishing & Roughness & Orientation\\[0.5ex]
\hline
S1 & Silica & $<$ 3 nm & (111) \\
S2 & diamond & $\geq$ 12 nm & (111)\\
S3 & no & $>$ 300 nm & (111)\\
\hline
\end{tabular}
\label{table:parameter}
\end{table}

The surface treatments are summarized in Table~\ref{table:parameter}. The measurements of the spin Seebeck effect (SSE) were performed in the following way. The samples were magnetized in the xy plane of the YIG crystal by an external magnetic field H, as shown in Fig.~\ref{fig:5}. 
To excite the spin waves an external heater generates a temperature gradient $\nabla T$ across the Pt/YIG stack where the temperature of heat sink is denoted as T. The thickness of the YIG (Pt bar) is 3 mm (15 nm) for all samples. Regarding the lateral dimension of the Pt bar, the length (width) varies from 5 mm-3 mm (2.5-1.5) with all samples having ratios 2:1. The surface of the YIG crystals was analyzed by atomic force microscopy (AFM) before deposition of the Pt film on top. The observed spin Seebeck signals show a small offset which we removed. The field at which 95$\%$ of the SSE signal saturates is defined as $H_{\text{sat}}^{\text{SSE}}$. The magnetization M of the YIG crystal with a dimension of 2~mm~x~1~mm was measured with a SQUID magnetometer.
\section{Results and discussion}
\begin{figure}[tbp]
 \centering
\includegraphics[width=0.5\textwidth,natwidth=310,natheight=342]{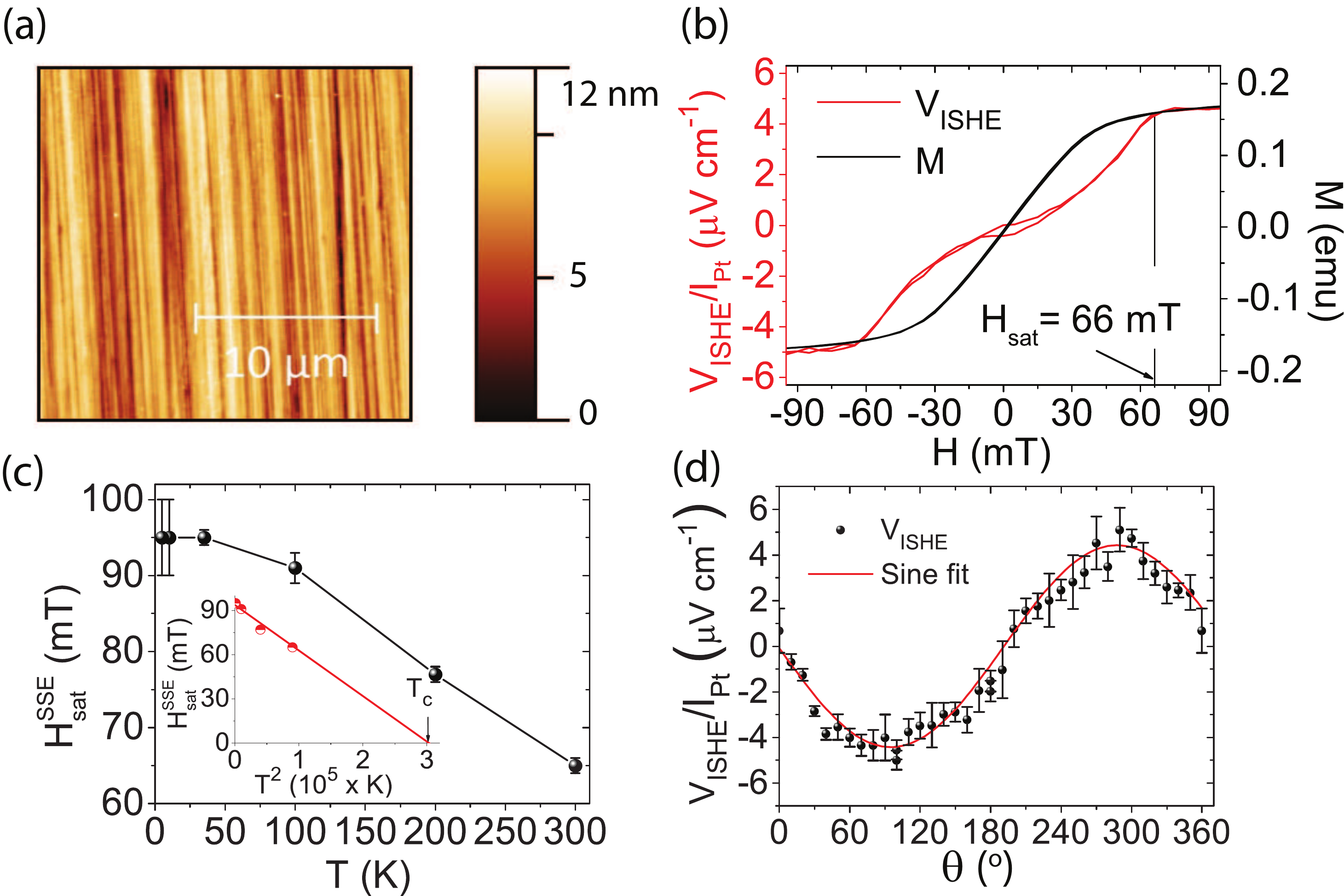}
\caption{\label{fig:1}
(a) AFM height image of a single crystal YIG surface for sample S2 (20 x 20 $\mu m^2$). (b) Comparison between the H dependence of $V_{\text{ISHE}}$ at $\Delta T$ = 3.6 K in sample S2 and the magnetization M of the YIG crystal. (c) Temperature dependence of $H_{\text{sat}}^{\text{SSE}}$. The inset shows $H_{\text{sat}}^{\text{SSE}}$ as a function of $T^\varepsilon$ where $\varepsilon$ = 2. (d) $V_{\text{ISHE}}$  as a function of the external magnetic field direction $\theta$ in the Pt/YIG system at a fixed magnetic field 80 mT.
}
\end{figure}

Fig.~\ref{fig:5}(c) shows the AFM height image of sample S1 with a surface roughness smaller than 3 nm. A distinct $V_{\text{ISHE}}$ signal appears and saturates around $\sim$ 55.3 mT, which is close to the field required to saturate the magnetization of the YIG crystal, as illustrated in Fig.~\ref{fig:5}(d).
Similarly, the YIG surface of sample S2 was analyzed by AFM. Fig.~\ref{fig:1}(a) shows that sample S2 has a surface roughness around $\sim$ 12 nm with strip-like trenches at the surface. A clear spin Seebeck response has been observed for sample S2 by changing the applied magnetic field H. The signal saturates at relatively higher values of H ($\sim$ 66.1 mT) compared to the magnetization of YIG as shown in Fig.~\ref{fig:1}b. 
In addition, we checked the magnetic field dependence of the spin Seebeck response at low-temperatures for sample S2, the temperature dependence of the $H_{\text{sat}}^{\text{SSE}}$ is given in Fig.~\ref{fig:1}(c). As the YIG crystal is a 3D isotropic ferrimagnet, the temperature dependence of the magnetic order parameter obeys a $T^2$ universality scaling \cite{Solt1962}. To understand the temperature dependence of $H_{\text{sat}}^{\text{SSE}}$, we fitted $H_{\text{sat}}^{\text{SSE}}$ at low temperatures by assuming $T_c$ = 553~K as shown in the inset of Fig.~\ref{fig:1}(c). The temperature dependence of $H_{\text{sat}}^{\text{SSE}}$ closely  obeys the $T^2$ universality  behavior of the order parameter of the YIG crystal with exponent $\varepsilon$ = 2. It suggests that the $H_{\text{sat}}^{\text{SSE}}$ directly depends on the order parameter of the YIG crystal. To confirm further the origin of the observed signal, H is rotated in the x-y plane. The $V_{\text{ISHE}}$ signal follows the expected sinusoidal dependence for a spin Seebeck signal, as shown in Fig.~\ref{fig:1}(d). 

\begin{figure}[hbt]
 \centering
\includegraphics[width=0.5\textwidth,natwidth=310,natheight=342]{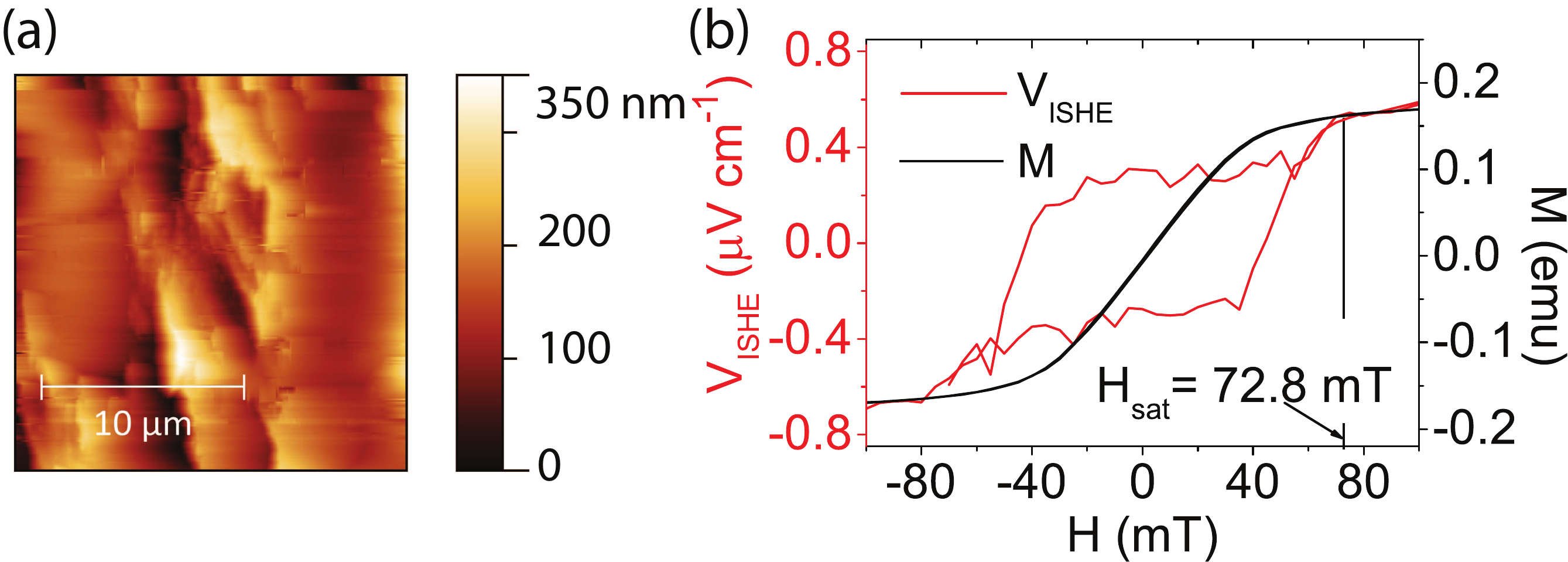} 
 \caption{\label{fig:4}
(a) AFM height image of the YIG surface for sample S3 (20 x 20 $\mu m^2$) and (b) a comparison between the H dependence of $V_{\text{ISHE}}$  at $\Delta$T = 7.5 K in sample S3 and the magnetization M of the YIG crystal.}
\end{figure}

Unlike the samples S1 and S2, sample S3 has a very large surface roughness ($>$ 300 nm) as shown in Fig.~\ref{fig:4}(a). Nevertheless, a clear spin Seebeck signal was observed as shown in Fig~\ref{fig:4}(b).

From equation~\ref{eq:V_ISHE}, it follows that the inverse spin Hall voltage $V_\text{{ISHE}}$ is proportional to the applied temperature gradient $\nabla T$ and the length of the Pt bar $l_{Pt}$. $V_{\text{ISHE}}$ increases by reducing the thickness of the Pt film $t_{Pt}$, for both the spin pumping \cite{Castel2012} and the SSE \cite{Saiga2014} experiments. Therefore to compare samples with different Pt thickness we can define a parameter C as follows \cite{Castel2012,Saiga2014,Nakayama2012}:
 
 \begin{equation} \label{eq:C parameter}
 C = \frac{1}{\textrm{tanh}[\dfrac{t_{Pt}}{2 \lambda_{Pt}}] \ \rho_{Pt} \ \dfrac{ l_{Pt}}{t_{Pt}}} \ \frac{V_{\text{ISHE}}}{\nabla T}
 \end{equation}

Here, $\nabla T$ is defined as the temperature difference across the Pt/YIG stack normalized with the thickness of the YIG crystal, $\rho_{Pt}$ is the resistivity of Pt and $\lambda_{Pt}$ is the spin diffusion length of Pt. In these experiments, unlike $\rho_{Pt}$, $\lambda_{Pt}$ cannot be measured directly therefore we assumed that it remains constant for different samples. Note that for all samples discussed here $t_{Pt}$~$>$~2~$\lambda_{Pt}$ (where $\lambda_{Pt}$ = $1.5~nm$ \citealp{Vlietstra_spinmix2013,flipse_observation_2013}) so the tanh$[\dfrac{t_{Pt}}{2 \lambda_{Pt}}]$ term is approximately equal to 1 leading to $V_\text{{ISHE}}$~$\propto$~$1/t_{Pt}$. Moreover, the C parameter is independent of the YIG thickness when the thickness is larger than the magnon mean free path and therefore it can be used as an indicator of changes in other parameters related to the interfacial mechanisms of the SSE.
\begin{table}[ht]
\small
\caption{Comparison of the resistance R of the Pt film, the  C parameter and the $H_{\text{sat}}^{\text{SSE}}$ for the SSE response in bulk single crystals and thin films}

\centering
\begin{tabular}{c c c c c c c}
\hline
\hline
 & & Bulk crystals & & & Thin films  \\[0.5ex]
\hline
& S1 & S2 & S3 & Ref.~\protect\cite{uchida2010} & Ref.~\protect\cite{flipse_observation_2013} \\[0.5ex]
\hline
R ($\Omega$) & 33.8 & 52.2 & 119 & - & - \\
\hline
C (10$^{-8}$V $\Omega^{-1}$ K$^{-1}$) & 0.917 &  1.369 & 0.043 & 0.554 & 1.105 \\
\hline
$H_{\text{sat}}^{\text{SSE}}$ (mT) & 55.3 & 66.1 & 72.8 & 40 & 2.5 \\

\hline
\end{tabular}
\label{table:comparison}
\end{table}

The resistance of the Pt film varies for the samples S1-S3, nevertheless all samples have similar resistance within an order of magnitude as shown in Table~\ref{table:comparison}. The observed change in the resistance is correlated with the roughness of the crystals, although we do not observe the same scaling for the SSE response. For example, the resistance of sample S2 is 50$\%$ higher than sample S1 whereas the SSE signal for sample S2 is only 30$\%$ higher than sample S1. Furthermore the resistance of  sample S3 is almost four times bigger than sample S1, however the SSE response actually follows the opposite trend, it is actually more than an order of magnitude lower than the response of the samples S1 and S2. Therefore, we establish that the dominant mechanism relevant for the observed differences in the SSE signal is not the resistivity of the NM films but the quality of the NM/FM interface. Sample S1 gives a C parameter that is comparable to the value reported for thin films and bulk crystals as shown in Table~\ref{table:comparison}. However, sample S2 shows 30$\%$ bigger and sample S3 shows more than an order of magnitude smaller value of the C parameter than sample S1. The observed variation in the value of the C parameter indicates the importance of mechanical treatment induced surface effects that we will discuss below.

\begin{figure}[tbp]
 \centering
\includegraphics[width=0.5\textwidth,natwidth=610,natheight=642]{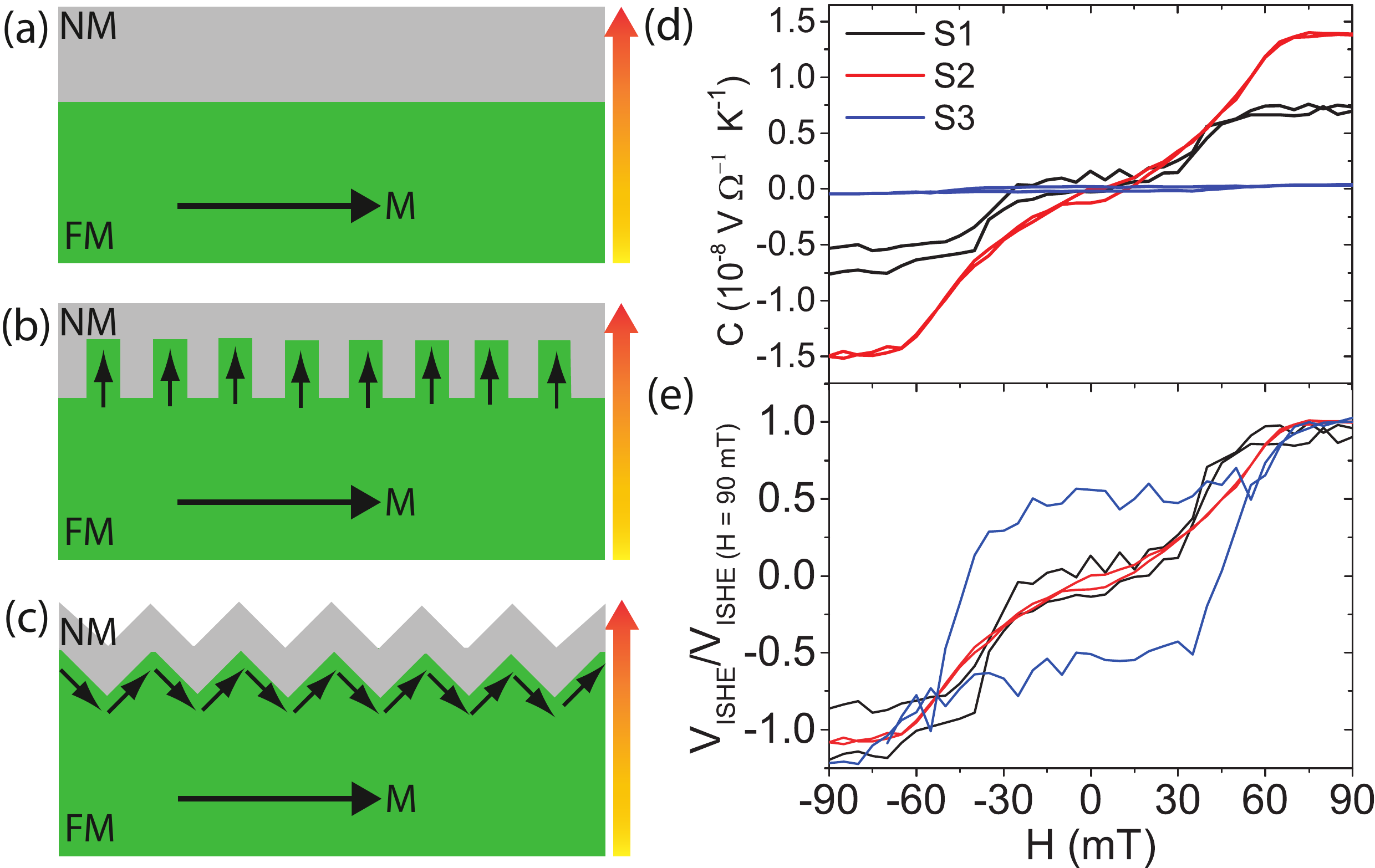}
\caption{\label{fig:8}
(a-c) A schematic illustration of the interface morphologies of the NM/FM system for different surface treatments of the FM where orange arrows represent $\nabla T$: (a) An atomically flat interface, (b)  an interface with a perpendicular anisotropy and (c) a rough interface. (d) Comparison between the magnitude of the C parameter and (e) comparison between the line profile of the SSE signal as a function of H for all samples. 
}
\end{figure}

Based on the experimental conditions listed in Table~\ref{table:parameter} and the results summarised in Table~\ref{table:comparison}, we propose a possible mechanism for our observations. Fig.~\ref{fig:8}(a-c) schematically illustrates possible interface morphologies and the surface magnetization for the NM/FM system, for different interface conditions between the NM film and the FM crystal. Fig.~\ref{fig:8}(a) represents the case for a NM film deposited on an atomically flat FM crystal. Here, the case for sample S1 corresponds to Fig.~\ref{fig:8}(a). Fig.~\ref{fig:8}(b) depicts a situation for a NM deposited on a flat FM crystal but having a small perpendicular anisotropy at the surface. The situation represented in Fig.~\ref{fig:8}(b) corresponds to the case for sample S2. The surface of the YIG crystal for sample S2 contains trenches due to polishing of the YIG crystal with coarse diamond particles as shown in Fig.~\ref{fig:1}(a). The trenches at the interface can induce strain or shape anisotropy resulting in a perpendicular anisotropy at the interface. The presence of a small perpendicular anisotropy at the interface would increase the $H_{\text{sat}}^{\text{SSE}}$ compared to the bulk magnetization of the YIG crystal, which has been clearly observed for sample S2 (see Fig.~\ref{fig:1}(b)). 

In addition, the magnitude of the SSE signal can also change if the mechanical polishing changes the atomic termination for the density of Fe atoms that are in direct contact with the Pt metal. If the density of Fe atoms at the surface is larger than the bulk of the YIG, the observed SSE signal would be larger \cite{UchidaFerrites2013,JiaSTT2011}. The increase in the SSE signal for sample S2 compared to sample S1 can be attributed to different chemical termination due to polishing with coarse diamond particles. Fig.~\ref{fig:8}(c) shows the case for a rough interface between the NM and the FM crystal which corresponds to the situation for sample S3. In case of sample S3, the lack of further mechanical treatment after cutting with a diamond saw leaves a  very rough surface of the YIG crystal. The $H_{\text{sat}}^{\text{SSE}}$ is around 72.8 mT for sample S3 as shown in Fig.~\ref{fig:4}(b). The increase in the value of $H_{\text{sat}}^{\text{SSE}}$ for sample S3 compared to the magnetization of YIG can be due to a non-uniform magnetization at the interface resulting from high surface roughness of the YIG crystal.

Fig.~\ref{fig:8}(d) gives a comparison of the magnitude of the SSE signal in terms of the C parameter (as defined in equation~\ref{eq:C parameter}) for samples with different mechanical treatments. The observed signal for sample S3 is smallest compared to other samples. This can be explained due to the increase of surface roughness \cite{Burrowes2012,JiaSTT2011} resulting in the small spin mixing conductance at the interface. The sample S1 has the lowest surface roughness, however the SSE signal observed for sample S2 is the largest compared to the samples S1 and S3 as shown in Fig.~\ref{fig:8}(d). Therefore, for the largest roughness of sample S3 we see a relation between roughness and the SSE signal, but not for the samples S1 and S2. Hence, the roughness is not the only parameter and this might be related to the more abrasive nature of the diamond particles leaving a different chemical termination at the interface.

To compare the line profile of the  $V_{\text{ISHE}}$ signal, in Fig.~\ref{fig:8}(e) the signals are normalized by their value at H = 90 mT, where they reach saturation. Fig.~\ref{fig:8}(e) shows that the line profile of the SSE signal changes with moving from soft silica to coarse diamond particle polishing. For the samples S1 and S2 the $V_{\text{ISHE}}$ is very small at zero applied field compared to the value measured at 90 mT. However, for sample S3 the $V_{\text{ISHE}}$ is almost 64$\%$ of the value measured at H = 90 mT. The value of  $H_{\text{sat}}^{\text{SSE}}$ is highest for sample S3 with the largest surface roughness and lowest for sample S1 with the smallest surface roughness. Therefore, the $H_{\text{sat}}^{\text{SSE}}$ directly correlates with the roughness of sample. The large deviation in the magnitude of SSE signal and the $H_{\text{sat}}^{\text{SSE}}$ in the YIG crystals with different surface treatments emphasizes the surface sensitivity of the spin Seebeck effect. Our results indicate that not only the surface roughness but actual atomic structures and chemical termination at the interface also play an important role in the SSE.

\section{Conclusions}
In conclusion, we have shown a strong dependence of the spin Seebeck signal on the interface condition of the Pt/YIG bilayer system. We observed a change of 18 mT in the saturation field of the SSE signal by changing the type of polishing. Furthermore we observe the change in the magnitude of the SSE signal for different samples. No definite relation has been found between the SSE response and the sample roughness. However, we observe a direct correlation between the $H_{\text{sat}}^{\text{SSE}}$ and the roughness of sample, as the former increases by moving from soft toward coarse particle polishing.  To understand the origin of the magnitude and change in the saturation field $H_{\text{sat}}^{\text{SSE}}$ for the observed SSE signal, due to different types of surface treatments, the crystal surfaces need to be investigated further in detail.

\begin{acknowledgments}
We would like to acknowledge J. Baas, H. Bonder, G.H. ten Brink, M. de Roosz and J. G. Holstein for technical assistance. This work is supported by the Foundation for Fundamental Research on Matter (FOM), the Netherlands Organisation for Scientific Research (NWO), Marie Curie ITN Spinicur NanoLab NL, and the Zernike Institute for Advanced Materials National Research Combination.
\end{acknowledgments}

%

\end{document}